\documentclass[useAMS,usenatbib]{mn2e}
\usepackage{natbib}

\usepackage[dvips]{graphics}
\usepackage{epsfig}
\usepackage{amsmath}
\usepackage{aas_macros}

\def\mearth{{\rm\,M_\oplus}}
\def\rearth{{\rm\,R_\oplus}}

\title[No universal minimum-mass extrasolar nebula]{No universal minimum-mass extrasolar nebula: Evidence against in-situ accretion of systems of hot super-Earths}
\author[Raymond \& Cossou]{Sean N. Raymond$^{1,2}$\thanks{E-mail: rayray.sean@gmail.com} and Christophe Cossou$^{1,2}$\\
$^{1}$Univ. Bordeaux, Laboratoire d'Astrophysique de Bordeaux, UMR 5804, F-33270, Floirac, France.\\
$^{2}$CNRS, Laboratoire d'Astrophysique de Bordeaux, UMR 5804, F-33270, Floirac, France
} 

\begin{document}

\date{Accepted to MNRAS Letters Jan. 15, 2014}

\pagerange{\pageref{firstpage}--\pageref{lastpage}} \pubyear{2013}

\maketitle

\label{firstpage}

\begin{abstract}
It has been proposed that the observed systems of hot super-Earths formed {\it in situ} from high-mass disks.  By fitting a disk profile to the entire population of Kepler planet candidates, Chiang \& Laughlin (2013) constructed a ``minimum-mass extrasolar nebula'' with surface density profile $\Sigma \propto r^{-1.6}$.  Here we use multiple-planet systems to show that it is inconsistent to assume a universal disk profile.  Systems with 3-6 low-mass planets (or planet candidates) produce a diversity of minimum-mass disks with surface density profiles ranging from $\Sigma \propto r^{-3.2}$ to $\Sigma \propto r^{0.5}$ (5th-95th percentile).  By simulating the transit detection of populations of synthetic planetary systems designed to match the properties of observed super-Earth systems, we show that a universal disk profile is statistically excluded at high confidence.  Rather, the underlying distribution of minimum-mass disks is characterized by a broad range of surface density slopes.  Models of gaseous disks can only explain a narrow range of slopes (roughly between $r^0$ and $r^{-1.5}$).  Yet accretion of terrestrial planets in a gas-free environment preserves the initial radial distribution of building blocks. The known systems of hot super-Earths must therefore not represent the structure of their parent gas disks and can not have predominantly formed {\it in situ}.  We instead interpret the diversity of disk slopes as the imprint of a process that re-arranged the solids relative to the gas in the inner parts of protoplanetary disks.  A plausible mechanism is inward type 1 migration of Mars- to Earth-mass planetary embryos, perhaps followed by a final assembly phase.
\end{abstract}


\section{Introduction}

The minimum-mass solar nebula (MMSN) is built by spreading the mass of the planets in concentric annuli then fitting a function to the distribution.  The original MMSN models produced surface density $\Sigma$ profiles that scaled with orbital radius $r$ as $\Sigma \propto r^{-1.5}$~\citep{weidenschilling77,hayashi85}.  Newer MMSN analyses have found solutions of $\Sigma \propto r^{-0.5}$~\citep{davis05} and $\Sigma \propto r^{-2.2}$~\citep{desch07}.  

The MMSN model almost certainly does not represent the initial conditions from which the planets formed.  It presumes that the current orbital architecture of the Solar System reflects the properties of its parent protoplanetary disk.  However, the giant planets are thought to have undergone multiple episodes of orbital migration; first during the gaseous disk phase~\citep{walsh11,pierens11} and later due to planetesimal-driven migration~\citep{malhotra95,tsiganis05}.  Models that do not invoke re-shuffling of the giant planets' orbits systematically fail to reproduce the inner and outer Solar System~\citep[e.g.,][]{levison08,raymond09c}.    

Thirty to fifty percent of main sequence FGKM stars host planets with masses $M \le 10-20 \mearth$, radii $R \le 2 \rearth$, and orbital periods $P \le 50-100$ days~\citep{howard10,howard12,mayor11,bonfils13,dong13,petigura13}.  These planets tend to be found in multiple systems in compact orbital configurations~\citep[e.g.][]{lovis11,lissauer11}.  The origin of these ``hot super-Earths'' is debated (see Table 1 in \citealt{raymond08a} as well as \citealt{gaidos07,raymond14}). The two most likely formation mechanisms are inward, type 1 migration of a population of planetary embryos~\citep{terquem07,ogihara09,cossou13b}, and {\em in situ} accretion of embryos from a massive disk~\citep{raymond08a,hansen12,hansen13,chiang13}.  The bulk density distribution of hot super-Earths should in principle be able to differentiate between the models~\citep{raymond08a}.  However, this requires a large number of precise radius and mass measurements of low-mass planets~\citep{weiss13,weiss13b} and a good understanding of the degeneracies between internal structure models~\citep{selsis07b,adams08}.  

Extra-solar planets can be used to create a minimum-mass {\em extrasolar} nebula (MMEN).  \cite{kuchner04} built an MMEN using 11 radial velocity-detected systems dominated by Jupiter-mass planets.  \cite{chiang13} built an MMEN using Kepler's small ($R < 5 \rearth$) planetary candidates by calculating a corresponding surface density for each candidate and performing a fit to the ensemble.  Both analyses produced steep density profiles: $\Sigma \propto r^{-2}$ for the gas giants and $\Sigma \propto r^{-1.6}$ for the super-Earths.

\cite{chiang13}'s MMEN is by construction the disk required to form hot super-Earths {\em in situ}. However, it is roughly a factor of seven more massive than the MMSN~\citep[][; see below]{raymond08a}, and it ties all hot super-Earths to a single disk or at least to a universal disk profile.  As we show in Section 2, the minimum-mass disks calculated from multiple-planet systems show a wide range of surface density slopes.  This broad distribution of slopes cannot be explained by selection effects or measurement errors (Section 3).  If hot super-Earths formed by {\em in situ} accretion the inferred disks must reflect the initial distribution of planetary embryos~\citep{raymond05b}.  However, many minimum-mass disks have very steep or flat profiles that cannot be reconciled with viscous disk models (Section 4).  We argue that this diversity does not represent the gas disk's true properties but rather is the product of a separation in the distributions of gas and solids in the inner parts of the disk, perhaps due to inward type 1 migration of planetary embryos.  

\section{Building minimum-mass disks}

Our sample of planets consists of the Kepler planet candidates~\citep{batalha13} and radial velocity systems downloaded via the Open Exoplanet Database~\citep{rein12}.  

We first reproduced the result of \cite{chiang13} using the Kepler candidates with $R<5 \rearth$ and $P < 100$ days.  We assigned planet masses by the simple relation~\citep{lissauer11b}:
\begin{equation}
\frac{M}{\mearth} = \left(\frac{R}{\rearth}\right)^{2.06}.
\end{equation}
We assigned each planet a surface density by simply calculating $\Sigma = \frac{M}{2 \pi a^2}$, where $a$ is the orbital radius.  The best-fit power-law was:
\begin{equation}
\Sigma(r) = 42.7 \,  \, \left(\frac{r}{\rm 1 AU}\right)^{-1.54} \, \rm{g \, cm^{-2}}.
\end{equation}
This is very close to the MMEN of \cite{chiang13}, which is reassuring.  


We next turned our attention to multiple-planet systems.  To include systems discovered by both radial velocity and transits, we only considered systems with at least three planets interior to 1 AU.  We restricted ourselves to systems containing only planets smaller than $5 \rearth$ or less massive than $30 \mearth$.  For Kepler candidates smaller than $1.5 \rearth$ we assumed $M \propto R^{3.7}$~\citep{valencia07} and for larger candidates we again assumed $M \propto R^{2.06}$~\citep[][; although we tested different M-R relations, see below]{lissauer11b}.  This effectively assumes an Earth-like composition for planets with $R < 1.5 \rearth$~\citep[consistent with observations; see][]{weiss13b}.  For radial velocity planets we used minimum masses.  

\begin{figure}
  \begin{center} \leavevmode \epsfxsize=8.5cm\epsfbox{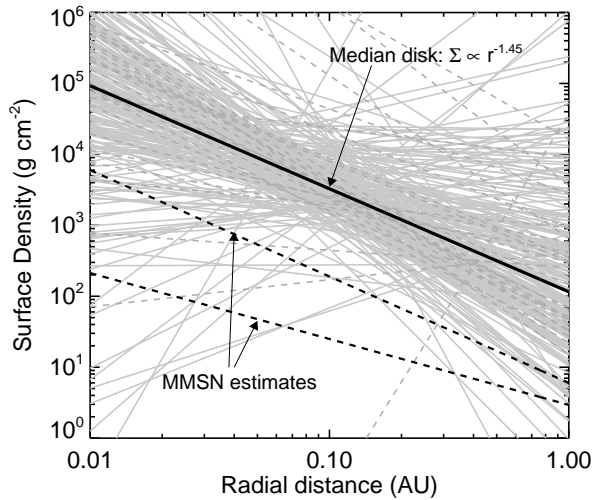}
    \caption[]{Diversity of minimum mass extrasolar nebulae.  Each grey curve is the minimum-mass disk inferred from a system of Kepler planet candidates (solid grey) or radial velocity planets (dashed grey).  The thick black curve is the median fit: $\Sigma \propto r^{-1.45}$.  The black dashed curves represents two estimates of the minimum-mass solar nebula (MMSN) built using just the terrestrial planets. } 
     \label{fig:mmsn}
    \end{center}
\end{figure}

For each system we calculated the width of the concentric annuli over which to spread the planets' masses as follows.  We first chose the midpoints between each pair of planets as the geometric mean of the two planets' orbital radii.  We then applied the same orbital radius ratios to extrapolate beyond the endpoints.  For instance, to extrapolate interior to planets 1 and 2 (with orbital radii $a_1$ and $a_2$, where $a_1 < a_2$), the inner edge of the innermost annulus is $r_{inner} = a_1\sqrt{a_1/a_2}$.  

Figure~\ref{fig:mmsn} shows the profiles of 191 minimum-mass disks (181 Kepler systems and 10 RV systems).  The median disk profile is
\begin{equation}
\Sigma(r) = 116 \,  \, \left(\frac{r}{\rm 1 AU}\right)^{-1.45} \, \rm{g \, cm^{-2}}.
\end{equation}
This is 2-3 times more massive than the universal MMEN we calculated above and that of \cite{chiang13}.  It contains $28 \mearth$ inside 1 AU and is 15-200 times more dense than the minimum-mass solar nebula (MMSN) in that range.  Fig.~\ref{fig:mmsn} includes two estimates for the MMSN.  The first was calculated as above considering only the Solar System's terrestrial planets: $\Sigma_{MMSN} = 2.94 \,  \left(\frac{r}{\rm 1 AU}\right)^{-0.93} \, \rm{g \, cm^{-2}}$.  The second is a profile shown by accretion simulations to reproduce the terrestrial planets~\citep{chambers01,raymond09c}: $\Sigma_{MMSN} = 6 \left(\frac{r}{\rm 1 AU}\right)^{-1.5} \, \rm{g \, cm^{-2}}$. 

\begin{figure}
  \begin{center} \leavevmode \epsfxsize=8.5cm\epsfbox{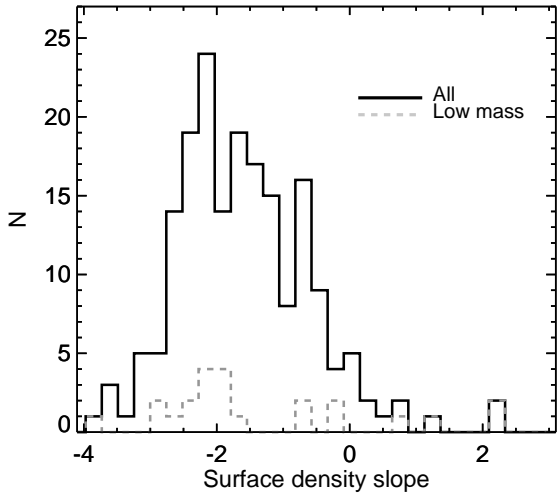}
    \caption[]{Histogram of the surface density slopes of minimum-mass disks (the grey curves from Fig.~\ref{fig:mmsn}.)  The binsize is 0.25 in width.  The dashed grey curve represents the distribution of slopes for systems only containing low-mass planets ($R_{max} \le 1.5 \rearth$ or $M_{max} \le10 \mearth$).}  \label{fig:hist}
    \end{center}
\end{figure}

There is a diversity of minimum-mass extrasolar nebulae (Fig.~\ref{fig:mmsn}). The surface density slopes range from -6.3 to +5.8.  19 disks have surface density slopes that increase with orbital distance, and another 19 have slopes of $r^{-2.5}$ or steeper.  Figure~\ref{fig:hist} shows the full distribution of disk slopes (excluding the widest outliers).  

Our results are only weakly sensitive to the chosen mass-radius relation.  We tested two $M$-$R$ relations in addition to our default, two-piece $M$-$R$ relation. First, we applied the $M \propto R^{2.06}$ relation of \cite{lissauer11b} to all planets regardless of mass.  Second, we applied the more complex, stellar flux-dependent mass-relation from \cite{weiss13}.  The MMEN slopes changed modestly between the different M-R relations: the median disk profile was $r^{-1.45}$ for both our fiducial and the \cite{weiss13} relations, and $r^{-1.31}$ for the \cite{lissauer11b} one.  K-S tests showed that the differences were not statistically significant.  


\section{Effect of sampling bias}
There exist biases in generating minimum-mass disks.  To address what underlying distribution could produce the inferred distribution of disks, we performed simple simulations of the detection of a population of planetary systems with a range of underlying characteristics.  

We generated synthetic planetary systems from disks with specified surface density profiles.  These systems were designed to roughly reproduce the underlying characteristics of the Kepler planetary candidate systems.  Each artificial system contained six planets.  The inner edge of the disk was assumed to be between 0.025 and 0.075 AU.  The period ratios among adjacent planets were drawn randomly between 1.3 and 3~\citep[see e.g.][]{rein12b}.  The planets' masses were calculated by assuming perfect accretion within a given region of the disk, with boundaries between adjacent planets chosen at the geometric means of their orbital radii.  The planet masses were in most cases between 1 and 15 $\mearth$.  The planets were given randomly-chosen inclinations of up to $5^\circ$\citep[][although we show the effect of the inclination distribution below]{tremaine12,fang12} and randomly-chosen longitudes of ascending node.  Each batch of simulations included $10^4$ systems.  

We simulated the transit detection of these artificial systems.  We assumed that the innermost planet was found to transit, and that the viewing angle was aligned with that planet's longitude of ascending node.  We then calculated whether each other planet would transit along that line of sight by testing whether its {\bf z} distance was between $-R_\star$ and $R_\star$, assuming a Sun-like host.  We did not transform the planetary masses into radii but rather simply attributed an error to the mass of each planet that was found to transit.  The detected mass was the true planet's mass times a factor that followed a log-normal distribution with $\pm 1 \sigma$ values of $2^{-1/2}$ and $2^{1/2}$.  This is simpler than more detailed studies~\citep{youdin11}, but it is sufficient for our purposes.  We performed the same fitting procedure on the simulated detections as for the real Kepler systems.  We only included systems with at least three planets detected within 1 AU.  This included cases where one or more planets were missed because of chance orbital alignment.  We compared the simulated detections with the Kepler systems.  

We first tested a universal disk profile with an $r^{-1.5}$ surface density slope.  With no uncertainties in measured masses and assuming perfectly coplanar systems, our algorithm retrieved the correct surface density slope, with a median of 1.5 and a standard deviation of 0.1.  For our default transit simulations with inclined orbits and mass errors the simulated systems had a median slope of 1.6 with a dispersion of 0.5.  The dispersion is determined in large part by the assumed mass uncertainty (see below).  The steeper slope comes from the fact that, when some planets are not found to transit, the annular area over which the detected planets' masses are spread to produce a surface density is increased such that the surface density is decreased.  Outer planets are more often missed than inner ones, which has a net steepening effect on the inferred profile, albeit with a large scatter.  We found evidence of this steepening in the data.  Kepler systems with packed orbital configurations -- defined such that the maximum orbital period ratio $P_{i+1}/P_i$ of adjacent planets is less than 2 -- have flatter minimum-mass disks than systems in which at least one pair of adjacent planets had $P_{i+1}/P_i > 2$.  The median surface density slope was -0.95 for packed systems and -1.77 for the more widely-spaced systems with enough dynamical space to harbor additional non-transiting planets.  The difference between the two was statistically significant ($p \sim 10^{-5}$ from a K-S test).  This steepening sets in at orbital inclinations of just $1-2^\circ$ and the distribution is independent of the underlying inclination distribution.  For large mutual inclinations a smaller number of systems have at least three transiting planets, but the inferred properties are essentially unaltered. 

\begin{figure}
  \begin{center} \leavevmode \epsfxsize=8.5cm\epsfbox{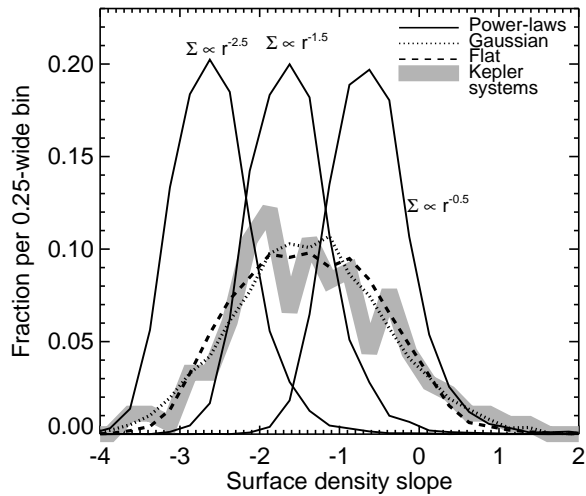}
    \caption[]{Probability distribution of the inferred surface density slopes for simulated observations of synthetic planetary systems.  The Kepler minimum-mass disks are shown with the thick gray line.  The solid black curves represent universal underlying power-law disks with profiles of $r^{-0.5}$, $r^{-1.5}$, and $r^{-2.5}$.  The other curves represent broad underlying distributions: a Gaussian with a mean radial exponent of -1.25 and standard deviation of 0.8 (dotted) and a flat distribution between zero and -2.5 (dashed).}  \label{fig:bias}
    \end{center}
\end{figure}

In-situ accretion from a universal disk profile is statistically inconsistent with the observations.  Although the $r^{-1.5}$ disks had a similar inferred median slope to the true Kepler systems, the distribution was far too narrow (Fig.~\ref{fig:bias}).  A K-S test showed a probability of $p \sim 10^{-10}$ that the two populations were drawn from the same underlying distribution.  The only way that we found to make the two distributions consistent was to dramatically increase the uncertainties in the inferred mass estimates to $\pm 1 \sigma$ values between $2^{-2.5}$ and $2^{2.5}$ (i.e., a factor of 30 within $\pm 1 \sigma$).  Such gigantic mass uncertainties are unrealistic.  All populations with assumed single power laws suffered the same problem of producing distributions that were far too narrow (Fig.~\ref{fig:bias}).  

The simulated detections reproduced the observed distribution if we assumed a broad underlying distribution.  Figure~\ref{fig:bias} shows two such examples that were consistent with observations (each with $p \sim 50\%$).  The first assumed a flat distribution of surface density slopes evenly spread between $x$=0 and $x$=-2.5.  The second assumed a Gaussian profile centered at $x$=-1.25 with a standard deviation of 0.8.  

In principle, a fraction of the observed systems could originate from in-situ accretion with a universal profile, as a wedge carved from the broad total distribution in Fig.~\ref{fig:bias}.  The maximum in-situ contribution is 30-50\% for our default assumptions but depends on the inferred slope distribution, which is dominated by errors in the inferred planet masses.  We interpret this upper limit as a significant overestimate because the remaining distribution would likely require multiple origins to explain its distorted bimodal shape.  It is far simpler to assume a broad distribution of underlying disks. 

There are other issues.  The radii of Kepler candidates are uncertain, especially for low-mass stars~\citep[e.g.][]{mann12}.  To ensure that this did not affect our results, we verified that our results are not a function of the stellar effective temperature.  We also note that many of the derived minimum-mass disks cannot be representative of the large-scale disk structure.  Disks with positive slopes contain thousands of Earth masses in solids if projected to AU radii, and disks with steep negative slopes contain very little total mass.   Finally,  we have only considered low-mass and small planets so our minimum-mass disks pertain only to the solid components of the disk.  Of course, if planets form {\it in situ} then the distribution of solids must reflect that of the gas~\citep{raymond05b}.  

\section{Implications for planet formation}

Gaseous protoplanetary disks are thought to evolve under the influence of internal viscosity~\citep[e.g.][]{armitage11}.  Viscous disk models produce surface density profiles with characteristic shapes.  They tend to zero at the inner edge and have a relatively shallow equilibrium slope extending out to close to the outer edge~\citep{lyndenbell74}.  For $\alpha$-viscosity disks~\citep{shakura73} the equilibrium profile is coupled to the temperature profile as:
\begin{equation}
T \Sigma \propto r^{-\frac{3}{2}}.
\end{equation}
A temperature profile dominated by stellar irradiation follows a $T \propto r^{-3/7}$ profile~\citep[e.g.][]{chiang97} with an equilibrium surface density profile of $\Sigma \propto r^{-15/14}$.  Disks in which the temperature is determined by viscous heating can have a range of profiles -- from $\Sigma \propto r^0$ (constant radial surface density) to $\Sigma \propto r^{-1.27}$ -- depending on the local opacity regime~\citep{dubrulle93,hure98,hersant01}.  A disk with a profile steeper that $\Sigma \propto r^{-3/2}$ must have a temperature that increases with radius, which is not physical in the inner parts of disks.  Alternately, steep profiles can be a result of an {\em outward} flow of gas in the outer parts of externally photo-evaporated disks~\citep{desch07}.  

Sub-millimeter observations of cold dust in the outskirts of protoplanetary disks consistently derive surface density profiles between $r^{-0.5}$ and $r^{-1}$(\citealt{mundy00,andrews09,andrews10}; although ~\citealt{isella09} measured a wider range of slopes that included positive exponents).  Interferometric measurements of molecular properties produce a similar range of density profiles but that extends to values as steep as $r^{-1.5}$~\citep{pietu07,guilloteau11}.  

If hot super-Earths formed {\it in situ} then minimum-mass disks must reflect the properties of the gaseous disks that spawned them.  The distribution of minimum-mass disks is much broader than the parameters for gaseous disks that can be easily explained by theory. The underlying distributions that are consistent with the Kepler disks are similarly broad (section 3).  How can we reconcile this diversity of minimum-mass disks with observations and theory?  

A simple solution is to invoke a separate evolution of solids and gas.  The planets from which minimum-mass disks are built would then not represent the structure of the full gas-dominated disk.  Differential solid-gas evolution can occur at m- or smaller sizes due to rapid aerodynamic drift~\citep{weidenschilling77b,boley13}.  Mars-mass or larger planetary embryos undergo type 1 migration via angular momentum exchanges with the protoplanetary disk~\citep{goldreich80,ward86}.  This migration can be directed inward or outward~\citep{kley08,paardekooper11} but is generally inward for embryos and planets with $M \le 5 \mearth$~\citep{bitsch13}.  

A leading model for hot super-Earth formation invokes accretion during the inward migration of a population of planetary embryos~\citep{terquem07,ogihara09,cossou13b}.  Migration is halted by the steep density transition at the inner disk edge~\citep{masset06}.  Embryos pile into chains of mean motion resonances.  Waves of inward-migrating embryos destabilize the resonant chains, trigger collisions and generate new resonant chains.  This increases the solid surface density in the inner parts of the disk and forms planets that follow a steeper profile than the underlying gas disk, although there is a range of inferred slopes (Cossou et al, in preparation).  

Resonant chains are a weakness of the type 1 migration model~\citep[but see][]{goldreich14}. The observed systems of hot super-Earth are rarely found in resonance~\citep{lissauer11b}.  The {\em in situ} accretion model can reproduce a number of features of the observed systems including their orbital spacings~\citep{hansen13,petrovich13}.  Late destabilization of resonant chains can act as a transition to the {\em in situ} accretion regime.  Destabilization can be triggered by the dissipation of the gaseous disk or chaotic dynamics~\citep{terquem07}.  

We thus favor a two-phase origin for hot super-Earth systems. First, inward migration of planetary embryos increases the solid surface density in the inner disk.  Second, the resonant chain is disrupted during a final phase of late-stage accretion.  This model reproduces current observations and predicts that a significant fraction (but not all) of hot super-Earths should be volatile-rich~\citep{raymond08a}.  

To conclude, we re-emphasize that minimum-mass nebulae -- both solar and extrasolar -- do not represent the underlying properties of the protoplanetary disks in which those planets formed.  A broad swath of studies have shown that planets and their building blocks undergo orbital migration at a range of size scales.  Minimum-mass nebulae have been sculpted by this migration.  

\vskip .1in 
\noindent 
We are grateful to referee Brad Hansen for a helpful review and to Franck Hersant and Franck Selsis for discussions.  We acknowledge the ANR for funding (project {\it MOJO}).  


\end{document}